\documentclass[12pt]{article}
\usepackage{graphicx}

\newcommand\pubnumber{SNSN-323-63}
\newcommand\pubdate{\today}

\def\Title#1{\begin{center} {\Large #1 } \end{center}}
\def\Author#1{\begin{center}{ \sc #1} \end{center}}
\def\Address#1{\begin{center}{ \it #1} \end{center}}

\newcommand\pubblock{\rightline{\begin{tabular}{l} \pubnumber\\
         \pubdate  \end{tabular}}}
\newenvironment{Abstract}{\begin{quotation}  }{\end{quotation}}
\newenvironment{Presented}{\begin{quotation} \begin{center}
             PRESENTED AT\end{center}\bigskip
      \begin{center}\begin{large}}{\end{large}\end{center} \end{quotation}}
\def\Acknowledgements{\bigskip  \bigskip \begin{center} \begin{large}
             \bf ACKNOWLEDGEMENTS \end{large}\end{center}}

\def\beq{\begin{equation}}
\def\eeq#1{\label{#1}\end{equation}}
\def\eeqn{\end{equation}}

\def\beqa{\begin{eqnarray}}
\def\eeqa#1{\label{#1}\end{eqnarray}}
\def\eeqan{\end{eqnarray}}

\let\bar=\overbar

\def\Dslash{\not{\hbox{\kern-4pt $D$}}}
\def\dslash{\not{\hbox{\kern-2pt $\del$}}}

\def\msb{{\bar{\ssstyle M \kern -1pt S}}}

\begin{document}
\begin{titlepage}
\pubblock

\vfill
\Title{The structure and analytic properties  \\
of the scattering amplitude  at LHC energies}
\vfill
\Author{{\underline{ Oleg V. Selyugin}$^1$} , Jean-Ren\'e Cudell$^2$\\[1ex]
 }
\Address{$^1$BLTPh, JINR,  Dubna, Russia, \\
 $^2$STAR Institute, Universit\'e de Liege, Belgium}
\vfill
\begin{Abstract}
Elastic proton-proton scattering is reviewed starting from the results of the LHC experiments conducted by the
TOTEM and ATLAS collaborations, and the HEGS model and a simple phenomenological  parametrisation are
compared with the new data on the differential
elastic proton-proton scattering
cross section, which detect a non-exponential behaviour of the differential cross sections in the first diffraction cone.
We consider the influence of various assumptions on the extraction of the elastic scattering parameters,
 and on the deduction of the total cross section.

\end{Abstract}
\vfill
\begin{Presented}
 EDS Blois 2017, \\
 Prague, Czech Republic, June 26-30,2017
\end{Presented}
\vfill
\end{titlepage}
\def\thefootnote{\fnsymbol{footnote}}
\setcounter{footnote}{0}
%
%
      Diffractive processes play an important role for our understanding of hadronic reactions at high
      energies. They result  from the exchange of a colour-singlet object called
      the  Pomeron that is the dominant contribution at the LHC, and which makes all diffractive processes
      grow with energy. The simplest way to describe it is to use a simple pole at $j=1+\epsilon$ \cite{Cudell-09},
      and unitarise it via an eikonal scheme.
      Elastic hadron scattering is the simplest diffractive reaction and it allows
      to investigate the basic properties of the  diffractive processes \cite{Rev-LHC}.
      The new data from the TOTEM  \cite{TOTEM} and ATLAS \cite{ATLAS}
         collaborations give very important information about this reaction.

         Unfortunately, the measurements of the elastic differential cross sections  at
        small momentum transfer and the subsequent derivation of  the total cross sections $\sigma_{tot}(s)$
        and of the values of $\rho(s,t=0)$
        $-$ the ratio of the real part to the imaginary part of the scattering amplitude $-$
        do not agree. The difference between the TOTEM and ATLAS results for the total cross section
        is $3$ mb at $7$ TeV
        and increases to $6$ mb at $8$ TeV. This is reminiscent of the conflicting
        Tevatron data for the total cross section at $\sqrt{s} = 1.8$ TeV.

        We shall explain here that the TOTEM data at $7$ TeV for the differential cross sections
        seem to be somewhat inconsistent with their determination of  $\sigma_{tot}(s)$ and $\rho(s,t=0)$
        \cite{Sel-NPA14}.
%
%
%
      To compare the data of the two collaborations, we need a model.
      We choose  the High Energy Generelazed  Structure model (HEGS), which satisfies the basic analytical properties
      of the $S$ matrix \cite{CudSel-16}, and describes
      quantitatively most existing experimental data in wide energy and momentum-transfer
      regions, including the Coulomb-hadron interference region.

      The model  describes very well
      the data at $\sqrt{s}=7$ TeV \cite{HEGS-0} and $8$ TeV \cite{HEGS-1} and gave predictions
      for $\sqrt{s}=13$ TeV \cite{HEGS-dm}. It has only a few free parameters and it embodies the
      main analytic requirements.  The real part of the amplitude can be obtained from crossing symmetry
      and the energy dependence comes from a single (unitarized) pomeron.
      When we compare the model with experimental data
       on the differential cross sections taking into account only statistical errors
       we find  that the ATLAS data is well described.
      However, the
      TOTEM data  are described only if one multiplies the results by normalization coefficients that depend on the dataset used,
      $n_{7 TeV}=0.97$ and  $n_{8 TeV}=0.91$, with other parameters
      parameters compatible with those from the fit to  ATLAS data.
      Hence, independently from the validity of the model, we find that the data
      from the TOTEM Collaboration lie slightly above those from ATLAS. It is likely
      that this is the cause of the difference between the values of the total cross section.
%
\begin{figure}[htb]
\centering
\includegraphics[height=2.in]{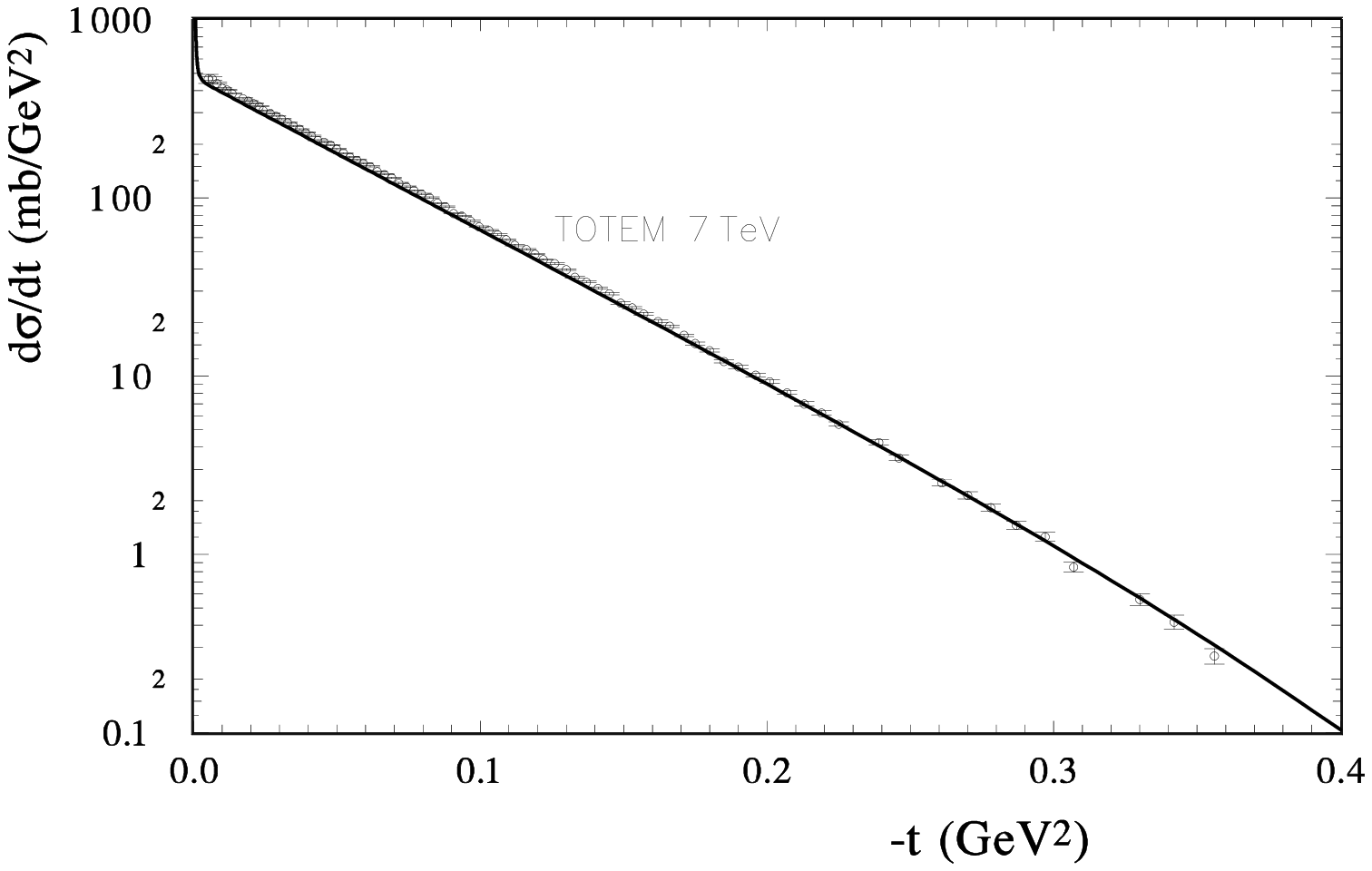}
\includegraphics[height=2.in]{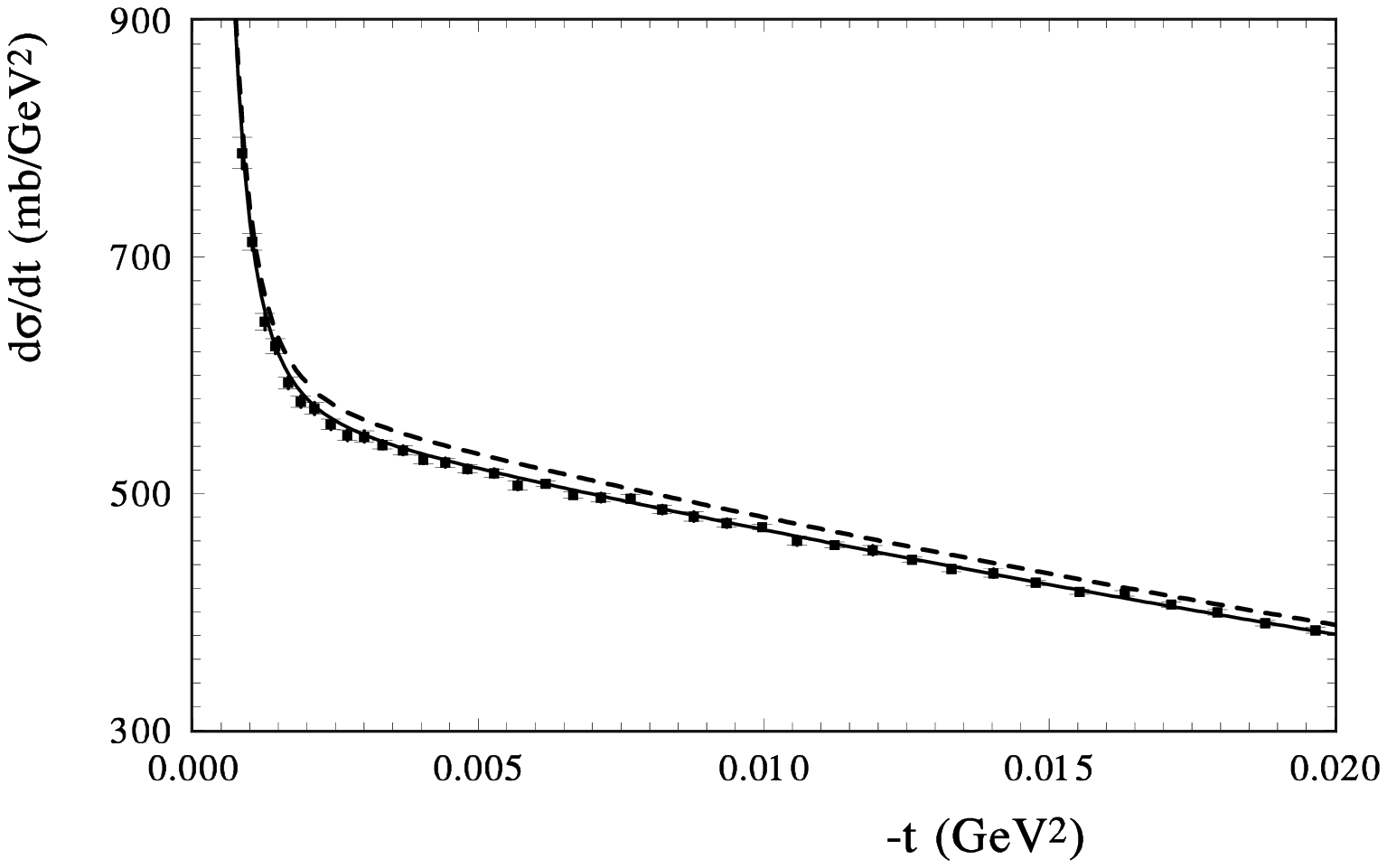}
\caption{Comparison of the differential cross sections obtained in the HEGS model
 with experimental data (the right-hand figure is for $\sqrt{s} = 7$ TeV and the left-hand one for  $\sqrt{s} = 13$ TeV and $\sqrt{s} = 14$ TeV
 and for the preliminary data at 13 TeV  \cite{T13s}.}
\label{fig:magnet}
\end{figure}
%

  We can also study the data on the differential cross sections
  though a simple phenomenological parametrisation  that includes an exponential form factor with a non-linear argument in $t$, and
  the possibility of additional normalization coefficients.
  The possibility of a non-exponential behaviour of the differential cross sections at small $t$ has a long
    history. One of the attempts to reproduce such a non-linear behaviour used
    the contribution of the 2-meson loop, which adds a term
    $\sqrt{(4 m_{\pi})^2-t} - 2 m_{\pi}$ to the pomeron slope,
    with $m_{\pi}$  the pion mass. At small $|t|$ such a term differs only very little from the standard slope $\alpha'=B_1t/2 $.
and one can use the approximation
    $$\sqrt{(4 m_{\pi})^2-t} - 2 m_{\pi} \approx  -{t\over 512 m_\pi} \left[128+8  {t\over m_{\pi}^2} \right]. $$
    The difference becomes significant only at $-t\geq 0.3$ GeV$^2$. This is the region that
     the original work tried to describe \cite{Gribov-Sl72}. For this term
    to matter at small $|t|$ one would need to reduce  $m_{\pi}$.

    Hence we take the scattering amplitude
    in the simple form
    $$ {\cal A}(s,t)  = \frac{i h s}{4 \pi} \log^{2}\left({s\over s_{0}}\right)
     \left( {s\over s_0}\right)^{{B_{1}\over2}t+{B_{2}\over 2} t^{2}} F_1^{2}(t) $$
    where the non-linear behaviour is described by the additional term $B_2 t^2/2$, $s_0=1$ GeV$^2$
    and $F_1(t)$ is the elastic form factor.
    We consider the three data sets of TOTEM  and the two sets of ATLAS.
    We first simultaneously analyse the data of both collaborations with only statistical errors
    and fix the additional normalization coefficient to $n_{i}=1.$
    In this case the $\chi^2$ is very large, as shown in the first column of Table 1.

    We then fit using systematic and statistic errors added in quadrature. In this case the $\chi^2$ substantially decreases, as seen in
    the second column of Table 1. If we then include the normalization coefficients in our fitting procedure
     as free parameters and consider again only statistical errors, we obtain a reasonable
     $\chi^2$, but the normalization coefficients are significantly different between TOTEM and ATLAS data.
     As in the HEGS model we see again that the data of the ATLAS collaborations are lower those of the TOTEM
     collaboration. Most remarkably, the obtained values for the total cross sections do not change much
     for all the considered variants. In all cases the value of $\sigma_{tot}(s)$ is close to that obtained by the
     ATLAS Collaboration.
\begin{table*} 
 \caption{ The results of the analysis of the experimental data on the differential cross sections,
  obtained by the TOTEM and ATLAS collaborations at $\sqrt{s}=7$ TeV and $\sqrt{s}=8$ TeV taking into account the additional
  normalization of the differential cross section  $n_{i}$ (the first column shows the results for
  statistical errors only, the second for  the  statistical and systematic errors added in quadrature, the third for statistical errors
  only, but with an extra normalisation factor for each dataset).
  }
\label{tab:1}
{\small\begin{center}
\tabcolsep12pt
\begin{tabular}{|c|c|c|c|}
\hline
                            &statistical   &  statistical+ & statistical+\\
                            &                  &  systematic & normalisation\\
\hline
$\chi^{2}$             &  48337     & 421              & 1812\\
 $h$ (GeV$^{-2}$)&  0.30        &0.30              &0.31\\
 $B_{1}$ (GeV$^{-2}$)&$0.55$&$0.55$&$0.58$\\
 $B_{2}$ (GeV$^{-4}$)&$-0.39$&$-0.39 $&$-0.26 $\\
 $\sigma_{tot} (7~\mathrm TeV)$ (mb)&95.3&95.1& 96.8\\
 $\sigma_{tot} (8~\mathrm TeV)$ (mb)&98.2& 98.0&99.7\\
 TOTEM $n_i$ at 7 TeV&&&0.97\\
 TOTEM $n_i$ at 8 TeV&&&0.95,0.94\\
ATLAS $n_i$ at 7 TeV&&&1.02\\
 ATLAS $n_i$ at 8 TeV&&&1.06\\\hline
\end{tabular}
\end{center}}
\end{table*} 
%

In conclusion, the new LHC data crucially constrain the  models of soft hadronic interactions.
We find that elastic scattering may reflect the generalized structure  of the hadron, from GPDs
which open a new way to connect elastic and inelastic interactions.
Also, the  standard eikonal approximation works well from  $\sqrt{s}=9$ GeV to  $8$ TeV.

Our calculations, based on the simplest phenomenological form of the scattering amplitude,
confirm the conclusions from the HEGS model that there is a difference in the normalization
of the differential cross sections in TOTEM and ATLAS data.
The new data show that the hadron interactions at large distances,
which reflects the properties of the differential  elastic cross sections
at small momentum transfer, can have complicated properties.
 This means that we need to examine carefully the fine structure of the diffraction peak
 to understand the effects of a non-exponential behaviour and maybe detect
 oscillations as the momentum transfer changes. This is tightly connected with the study of possible
 contributions from the hard Pomeron and the Odderon. All these questions may have an effect on the
 determination of the total cross section and on the extraction of the ratio of the real part to the imaginary part
  of the scattering amplitude. New experimental data at $13$ and $14 $ TeV, measured with high
   accuracy, will shed new light on these issues.
\newpage
\Acknowledgements
O.V.S.  would like to thank  
Prof. Marek Tasevsky for the invitation
and support 
 gratefully acknowledges  the financial support
  from FRNS  and would like to  thank the  University of Li\`{e}ge
  where part of this work was done.

\end{document}